%% file: 00-main.tex
\documentclass[conference]{IEEEtran}

\input{01-packages}

\begin{document}

\input{02-top-matter} 

\maketitle

\input{03-abstract.tex}

\input{10-intro}
\input{20-data}

\input{25-requirements.tex}
\input{30-metricspace.tex}

\input{40-experiments}

\input{50-conclusion}

\input{90-acks} 
\small
\bibliographystyle{IEEEtran}
\bibliography{refs}

\end{document}

%% file: 01-packages.tex
\usepackage[dvipsnames]{xcolor} 
\usepackage{wrapfig}
\usepackage{booktabs}
\usepackage[bookmarks=false]{hyperref}
\usepackage{amsmath, amssymb, amsfonts, amscd} 
\usepackage{amsthm}
\usepackage[justification=raggedright]{subcaption}
\usepackage{courier} 
\usepackage{graphicx}
\graphicspath{ {images/} } 
\usepackage{multicol}
\usepackage{lipsum} 
\usepackage[numbers,sort&compress,square]{natbib} 
\usepackage{enumitem} 
\usepackage{todonotes}

\usepackage{listings}
\usepackage{inconsolata}

\usepackage{threeparttable}
\usepackage{array}
\usepackage{multirow}

\newcommand\norm[1]{\left\lVert#1\right\rVert}

\colorlet{punct}{red!60!black}
\definecolor{delim}{RGB}{20,105,176}
\colorlet{numb}{magenta!60!black}

\lstdefinelanguage{json}{
    basicstyle=\fontsize{6}{3}\selectfont\ttfamily,
    keywords={BaseFileName, BaseFileExtn, CreatorProc, NewProc, TokenElevation, CmdLine,   FileDesc, Time},
    keywordstyle=\ttfamily\bfseries,
    showstringspaces=false,
    breaklines=true,
    literate=
     *{:}{{{\color{punct}{:}}}}{1}
      {,}{{{\color{punct}{,}}}}{1}
      {\{}{{{\color{delim}{\{}}}}{1}
      {\}}{{{\color{delim}{\}}}}}{1}
      {[}{{{\color{delim}{[}}}}{1}
      {]}{{{\color{delim}{]}}}}{1},
}

\theoremstyle{plain}
\newtheorem{theorem}{Theorem}[section]

\newtheorem{corollary}[theorem]{Corollary}
\newtheorem{proposition}[theorem]{Proposition}  

\theoremstyle{definition}
\newtheorem{definition}[theorem]{Definition}

\theoremstyle{remark}

\newcommand{\thm}{\begin{theorem}}
\newcommand{\ethm}{\end{theorem}}
\newcommand{\defn}{\begin{definition}}
\newcommand{\edefn}{\end{definition}}
\newcommand{\eprop}{\end{proposition}}
\newcommand{\prop}{\begin{proposition}}
\newcommand{\cor}{\begin{corollary}}
\newcommand{\ecor}{\end{corollary}}

\IEEEoverridecommandlockouts 

\usepackage{outlines}
\usepackage{dblfloatfix}

%% file: 02-top-matter.tex
\title{\LARGE{Defining a Metric Space of Host Logs and Operational Use Cases}
\thanks{\footnotesize{This manuscript has been authored by UT-Battelle, LLC under Contract No. DE-AC05-00OR22725 with the U.S. Department of Energy. The United States Government retains and the publisher, by accepting the article for publication, acknowledges that the United States Government retains a non-exclusive, paid-up, irrevocable, world-wide license to publish or reproduce the published form of this manuscript, or allow others to do so, for United States Government purposes. The Department of Energy will provide public access to these results of federally sponsored research in accordance with the DOE Public Access Plan (\url{http://energy.gov/downloads/doe-public-access-plan}).} }
}

\author{
\IEEEauthorblockN{Miki E. Verma, Robert A. Bridges}\\
\IEEEauthorblockA{Cyber \& Applied Data Analytics Division, Oak Ridge National Laboratory, Oak Ridge, TN\\
\{vermake, bridgesra\}@ornl.gov}}

%% file: 03-abstract.tex
\begin{abstract}
Host logs, in particular, Windows Event Logs, are a valuable source of information often collected by security operation centers (SOCs). 
The semi-structured nature of host logs inhibits automated analytics, and while manual analysis is common, the sheer volume makes manual inspection of all logs impossible. 
Although many powerful algorithms for analyzing time-series and sequential data exist, utilization of such algorithms for most cyber security applications is either infeasible or requires tailored, research-intensive preparations. 
In particular,  basic mathematic and algorithmic developments for 
providing 
a generalized, meaningful similarity metric on system logs is needed to bridge the gap between many existing sequential data mining methods and this currently available but under-utilized data source.
In this paper, we provide a rigorous definition of a metric product space on Windows Event Logs, 
providing an embedding that allows for the application of established machine learning and time-series analysis methods. 
We then demonstrate the utility and flexibility of this embedding with multiple use-cases on real data: 
(1) comparing known infected to new host log streams for attack detection and forensics, 
(2) collapsing similar streams of logs into semantically-meaningful groups (by user, by role), thereby reducing the quantity of data but not the content,
(3) clustering logs as well as short sequences of logs to identify and visualize user behaviors and background processes over time. 
Overall, we provide a metric space framework for general host logs and log sequences that respects semantic similarity and facilitates a wide variety of data science analytics to these logs without data-specific preparations for each. 
\end{abstract}

%% file: 10-intro.tex
\section{Introduction}
\label{sec:intro}
Recently, our research team has interacted with six security operation centers (SOCs): touring centers, interviewing analysts to understand their work, data, and needs, and working closely with operators for research and development projects.  
Security operations now have widespread collection and query capabilities for host-level data, and cyber operators continually monitor the alerts, warnings, and other information reported by the anti-virus (AV) and system logging capabilities of each of the IPs in their network.
While cyber operators regularly manage these large quantities of time-series data, most investigations and decisions require manual analysis, and thus, much of the potential of these data-rich, information-impoverished sources are not leveraged. 

One particular, readily available, but obtuse source of data are host system logs. These detail an extensive array of events on a host (e.g. file executions, user command prompts, socket creations), information that is invaluable for cyber operations tasks such as triage and forensics. 
A flexible log data structure allows for encoding this broad range of information, but the lack of uniformity and mixed data types are prohibitive for data analysis. 
While powerful mathematics and resulting algorithms for analysis of time-varying data exist, such algorithms are often impotent in cyber applications, or require time-consuming, data-specific preparations before application. 
Therefore, there is a need for an embedding of logs into a more manageable format while preserving semantics and context. 
Such a representation would bridge the gap between currently available streams of system logs and many mature, tested algorithms for structured, sequential data. 

\input{12-contributions.tex}

\input{14-related-work.tex}

%% file: 12-contributions.tex
\subsection{Contributions}
Our approach is to define a metric space on system logs, which provides a computable distance for both single logs and time-varying sequences of logs. 
The formulation is general enough to be applied to diverse data streams with minimal configuration.
Furthermore, the distance helps to preserve semantic similarity across logs/log streams, permitting more automated and meaningful analytics; for example, softening the brittle signature-based detection methods that currently dominate malware file conviction technologies. 

Armed with this distance measure, we open the door to a  number of existing machine learning and time-series analysis methods which offer solutions to the three of the biggest issues for SOCs:  
(1) a huge volume of data by facilitating dimension reduction and clustering techniques; 
(2) finding unusual data and spotting patterns via anomaly detection and  correlations methods; 
(3) incorporating context in which logs occur using time-series analysis to view logs in their natural sequence.

By working with operators, we  gained valuable insight that contributed to this research: 
determining a priori which aspects of the logs were most useful, 
learning what general patterns operators seek and what these patterns indicate, 
verifying that our embedding was still semantically meaningful to operators (that  is, their own intuition would not be lost when looking at the transformed logs), and finally, designing some experiments to show the utility of our proposed embedding, especially for security analysts. 
The operators also provided us with a set of Windows Event Logs (systems logs native to Windows OS) from a number of different IPs, including streams of data collected during multiple instances of a ransomware attack (CryptoWall 3.0).

In Sec. \ref{sec:data} we give details on Windows Event Logs, discussing some of the current uses and challenges with this data, and in Sec. \ref{sec:requirements} we enumerate requirements for our metric space and motivate our approach with a series of examples. 
In Sec. \ref{sec:metricspace}, we rigorously define our metric space on host logs. 
We first define a generalized metric space for log entries, providing a quantifiable and meaningful similarity measure between standardized log representations and permitting clustering, dimension reduction, and other parametric machine learning techniques. 
We further extend this metric to time-ordered sequences of logs produced by a single IP, facilitating the application time series analysis methods and allowing for the incorporation of context through temporal aspects of the data. 
In Sec. \ref{sec:experiments} we present  preliminary experiments that exemplify the utility and flexibility of our embedding\textemdash polymorphic malware detection by comparing streams of data during an attack to new ambient data, user IP and role classification by grouping similar substreams, clustering logs to visualize user activity per hour, and clustering short sequences of logs to visualize a host's logs throughout a day\textemdash followed by conclusions,  Sec. \ref{sec:Conclusion}.


%% file: 14-related-work.tex
\subsection{Related Works}

To our knowledge no previous research work has attempted to define a metric space for host log streams. There have been some efforts to develop more systematic ways of using Windows Event Logs to detect malicious events.

Anthony~\cite{anthony2013detecting} details a number of strategies and provides insight for hunting threats in Event Log data. He enumerates several standard patterns that may indicate various stages of compromise (i.e persistence, privilege escalation, lateral movement, etc.), and suggests standard sets of filters to use when hunting different types of threats.

Dwyer \& Truta~\cite{dwyer2013finding} propose an automated way of processing and detecting anomalies in streams of Windows Event Logs. They detail the process of filtering and ingesting logs into SQL database, and introduce standard-deviation based alerting system hinged on basic summary statistics. 

Aharon et. al.~\cite{aharon2009graph} develop two novel methods for automated processing and pattern recognition in semi-structured text-based event logs: a sequential text clustering algorithm for transforming logs into different event types, and the PARIS (Principle Atom Recognition in Sets) algorithm to identify and isolate co-occurring processes in a single data stream. 
They implement these algorithms on real log data and then  apply the results to tasks such as characterizing system behavior and efficiently searching and visualizing logs.

Berlin et. al.~\cite{berlin2015malicious} build a set of features from Windows audit logs and apply a linear classification model towards a robust intrusion detection system. Testing their approach on a wide range of malicious and benign software samples, they report a 78\% detection rate on malware missed by AVs.

Chen \& Bridges~\cite{chen2017automated} introduce a method to automatically extract ransomware behavior from noisy system logs. 
The technique ignores sequential context by building ``bag-of-words'' features, that is, counts of events in a time-series of logs. 
Taking input as log streams with and without malware, the method extracts the features caused by malware. 
Using WannaCry and variants they show the method is robust to polymorphism that tricks 64/64 AV checked by VirusTotal (\url{www.virustotal.com}).

A separate but related area of research known as ``log reduction'' attempts to significantly decrease the number of system audit logs without information loss in order to simplify the process of forensic analysis, e.g.,   \cite{hossain18dependence, xu2016high}.

%% file: 20-data.tex
\section{Event Log Analysis: Data \& Challenges}
\label{sec:data}
Developments in this paper progress on a private dataset of real Windows Event logs donated to us from a cooperating SOC. 
The logs are from  12 hosts, with contiguous captures varying between one to eleven days.  
This test set is slightly over  0.5GB comprising 88 IP-days for an average of about 6MB / IP / day. 
To put this in perspective, medium-sized operations will collect similar host logs for approximately 10K IPs continuously. 
Our operators have reported total host data collection (across a variety of data sources) of 300MB to 10TB per day depending on organization size and configurations of what is collected. 

Importantly, for our experiments, annotations found by manual investigations of the operators were given,  including the role of the user (e.g., IT professional, administrative assistant) and labeled attacks\textemdash four captures contain logs leading up to and during an infection with CryptoWall 3.0, a prominent ransomware from 2015 \cite{cryptowall, cryptowall-1, cryptowall-2}.  

While we focus on Windows Event Logs, we note that our formulation is sufficiently general to apply to a wide variety of host logs. 

\subsection{Windows Event Logs \& Limitations} 
\begin{figure}[t]
\centering
 \includegraphics[width=0.4985\textwidth]{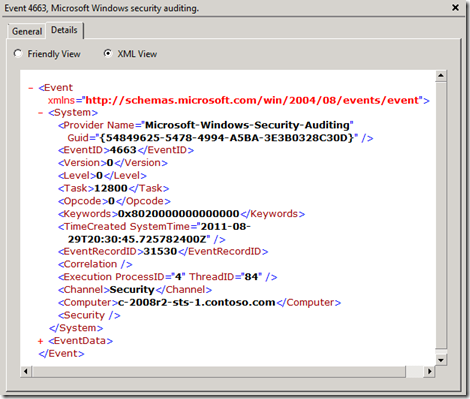}
\caption{Windows Event Log Metadata viewed in Event Viewer Interface as an XML. Photo credit Microsoft Technet Blog \url{https://goo.gl/HC1enm}
}
\label{fig: eventxml}
\end{figure}

Windows Event Logs are semi-structured records that detail nearly all software and hardware related events that occur on a Windows machine. 
They are generally comprised of three types of logs: application, system, and security~\cite{windowseventlog}.
These events can be viewed in the native Event Viewer\textemdash see Fig.~\ref{fig: eventxml} for an example log in the native interface\textemdash yet more commonly SOCs configure hosts to filter a subset of the logs and automatically forward them to a security information and event management system (SIEM). 

The non-relational data structure (e.g. XML, JSON) of the event metadata permits flexibility in logs and facilitates easy viewing and filtering by SOC operators. 
Hence, host logs are a primary source used to trace attacks and find patterns. 
For example, see our previous works~\cite{chen2017automated, bridges2018forming} describing manual and automated processes for malware analysis through log investigations. 
However, the often nested structure and non-uniform keys / values are not well-suited to algorithms that attempt to accomplish these same tasks in an automated way. 
As a result, many of the analytic and alerting tools SOCs have to analyze event logs provide only very basic statistics and brittle rule-based detection.  
More complex analysis and nuanced comparisons between logs are  left to the discretion of these domain experts. 

One of the main impediments in applying computational methods to event log data is the inability to automate these types of nuanced comparisons. Currently, the only way to compare logs or log attributes is by a simple binary classification \textemdash either two logs/attributes are identical, or they are different. 
This limitation reflected in the way that current tools operate on logs, specifically:  
basic statistical analysis relies on the few quantifiable aspects of these logs, e.g., number of logs with a new Windows Explorer process created in the last hour; 
querying involves grouping logs that share identical values for the same key (which then takes them out of the chronological context in which they occurred); 
rule-based detection is based on a determination as to whether a new log's attribute satisfy a rigid heuristic, e.g., are the same as a previous log.  

\textit{Overall, there is a need for a representation of host logs that permits a semantically meaningful similarity measure that is more flexible than a simple binary classification, and that exploits the temporal aspects of the data, incorporating the context in which the log appears.  
}

%% file: 25-requirements.tex
\section{Metric Space Requirements} 
\label{sec:requirements}
In this section, we outline three requirements for our similarity measure. 
For each, we provide motivation from specific examples of host logs. We note that examples used in this section are not contrived. 
They are anonymized real logs that have undergone preprocessing that is part of our method (see Sec. \ref{sec:defset}), specifically, nested logs are flattened, and only a select subset of the the more-than-200 keys are shown. These examples concisely illustrate our requirements, but do not fully capture the sparse but high-dimensional nature of our data.

\textit{Requirement 1}: Our first requirement is that each log can be compared attribute-wise, and that our measure can encode degrees of similarity that match the logs meanings. Shown in Fig. \ref{fig:cryptotempfilelogs} are two logs from different hosts during the initial portion of the Cryptowall attack in which a temporary file is installed with a random filename. 
While basic visual inspection reveals that these logs are extremely similar, without such a similarity metric, these logs would be considered strictly different. 
The slight difference in these logs, namely the randomized file name, exhibits Cryptowall's polymorphism; hence, not discriminating on this discrepancy is vital for identifying the pattern Cryptowall leaves in logs (and the perennial Achilles' heel for signature-based detection). 
While this is a particularly critical example, more common examples are easy to contrive, e.g., two logs vary by only the timestamp, IP, and directory from which a shell command is run\textemdash the primary action, that the same command is run in both, should be captured by the similarity score. 
\begin{figure}[h]
\noindent\begin{minipage}{.24\textwidth}
\begin{lstlisting}[language=json]
{ CmdLine: C:\Users\hostY\AppData\Local\Temp\Low\8D6.tmp
  CreatorProc: iexplore,
  NewProc: C:\Users\hostY\AppData\Local\Temp\Low\8D6.tmp
  TokenElevation: 3 }
\end{lstlisting}
\end{minipage}\hfill
\begin{minipage}{.24\textwidth}
\begin{lstlisting}[language=json]
{ CmdLine: C:\Users\hostX\AppData\Local\Temp\Low\4328.tmp
  CreatorProc: iexplore,
  NewProc: C:\Users\hostX\AppData\Local\Temp\Low\4328.tmp
  TokenElevation: 3 }
\end{lstlisting}
\end{minipage}
\caption{Examples of two truncated logs from two different IPs during Cryptowall 3.0 attack depicted, illustrating Requirement 1. While clearly similar, without a semantic-preserving metric, the two logs will be considered completely different. We note that  \texttt{.tmp} files with random names exhibit the polymorphism of Cryptowall 3.0. }
\label{fig:cryptotempfilelogs}
\end{figure}

\textit{Requirement 2}: The second requirement for our embedding is that it allows us to create an average representation of a group of logs (e.g., the centroid of a cluster). 
Note that this implies we can use distance to this mean  to quantify how similar a single log is to a group. 
We additionally require that these average representations retain semantic meaning and remain interpretable. 
Fig. \ref{fig:explorerlogs} shows three new process logs with the same new process (Windows Explorer), but that vary in the creator process and the executable's directory.
The left and center logs typify the normal (non-attack) Windows Explorer process logs.
These show a new Windows Explorer process being launched by a startup process and a second Windows Explorer process respectively (not uncommon). 
Yet, the third shows that a temporary file named \texttt{4328} opened Windows Explorer\textemdash  anomalous and seen during a Cryptotowall attack \cite{cryptowall}. 
Ideally, our embedding would allow us to compute an average over all logs with Windows Explorer as a new process (finding a centroid between the observed creator processes),
quantify how much variation there is within these logs, 
and measure the anomalousness of any given new Windows Explorer log. 

\begin{figure}[h]
\noindent\begin{minipage}{.16\textwidth}
\begin{lstlisting}[language=json]
{ CmdLine: C:\Windows\explorer.exe
  CreatorProc: userinit,
  FileDesc: Windows Explorer,
  NewProc:C:\Windows\explorer.exe
  TokenElevation: 3 }
\end{lstlisting}
\end{minipage}\hfill
\begin{minipage}{.16\textwidth}
\begin{lstlisting}[language=json]
{ CmdLine: C:\Windows\Explorer.exe
  CreatorProc: explorer,
  FileDesc: Windows Explorer,
  NewProc: C:\Windows\explorer.exe
  TokenElevation: 3 }
\end{lstlisting}
\end{minipage}
\begin{minipage}{.16\textwidth}
\begin{lstlisting}[language=json]
{ CreatorProc: 4328,
  FileDesc: Windows Explorer,
  NewProc: C:\Windows\SysWOW64\explorer.exe
  TokenElevation: 3 }
\end{lstlisting}
\end{minipage}
\caption{Examples of three truncated logs depicted, each reporting  Windows Explorer is a new process spawned by a different creator process.  
Left, Center logs are frequently seen in ambient data. 
Right log is only seen during Cryptowall 3.0 attack. 
These illustrate motivation for Requirement 2\textemdash a similarity metric should allow  a semantically-meaningful average representation, which could be used in this case 
to identify that the rightmost log is anomalous (unusually far from the average).}
\label{fig:explorerlogs}
\end{figure}

\textit{Requirement 3}: Finally, we wish our embedding to be able to incorporate the context in which the log appears, both in time and in sequence, as most distinct incidents occur over a sequence, and cannot be fully understood from a single log. 
Fig. \ref{fig:flashplayerlog} shows the attack vector log that is common to all of the attack instances seen in our dataset\textemdash a Flash upgrade that was presumably harboring the malware. 
However, this same log is also seen in ambient data, presumably for legitimate Flash upgrades. 
Hence, analyzing this log independently will not yield important information, namely, that it precipitates every sequence of events (e.g., anomalous logs discussed above) leading to the encryption of files by Cryptowall.  
Consequently, our metric should extend to quantify the similarity between time ordered sequences of logs. 
\begin{figure}[!h]
\begin{lstlisting}[language=json]
{ CmdLine: C:\Windows\system32\Macromed\Flash\FlashUtil64_19_0_0_185_ActiveX.exe -Embedding
  CreatorProc: svchost,
  FileDesc: Adobe\u00ae Flash\u00ae Player Installer/Uninstaller 19.0 r0,
  NewProc:C:\Windows\System32\Macromed\Flash\FlashUtil64_19_0_0_185_ActiveX.exe
  TokenElevation: 3 }
\end{lstlisting}
\caption{Example of truncated log containing the attack vector of all Cryptowall 3.0 attack instances in our in our dataset is depicted. This log is also seen in ambient data, so viewing it out of context is insufficient. This motivates Requirement 3---a similarity measure between sequences of logs in order to view the log in context.}
\label{fig:flashplayerlog}
\end{figure}

        
        

        


%% file: 30-metricspace.tex
\section{Defining a Metric Space}
\label{sec:metricspace}
Here we provide the metric space definitions and foundations for  logs and log streams. 
First we recall the definition of a metric space.

\defn [Metric Space]
	\label{defn:metricspace}
    A {\it metric space} is a set $X$,
    with a metric $d: X \times X \mapsto \mathbb{R}$, a function that maps every pair of elements to a distance such that for all $x,y,z \in X$, the following properties are satisfied:
    
    \begin{enumerate}[label=\bfseries\arabic{*}.]
    \item (Positivity) $d(x,y) \geq 0$ and  $d(x,y) = 0 \Longleftrightarrow x=y$
    \item (Symmetry) $d(x,y) = d(y,x)$
    \item (Triangle Inequality) $d(x,y) + d(y,z) \geq d(x,z)$
\end{enumerate}
\edefn 

The rest of the section gives rigorous construction of the metric space on logs.  
As a guide, we enumerate the process as the following five steps. 

\begin{enumerate}
\item {\it Un-nest: } Flatten nested fields in the raw logs. 
\item {\it Standardize records:} Make all logs have a uniform set of keys. Fix a subset of desired keys and their value types (a schema). For each record, discard other keys, and if necessary add the desired keys  with a null value. 
\item {\it Define a  metric per type: } Define a metric $d_i$ on each value type (numerical, categorical, string). 
\item {\it Extend to a metric for records: } Extend the component-wise metrics to a  distance between records using $\ell^2$ norm. 
\item {\it Extend to a metric for sequences of records: } Use $\ell^1$ norm on sequences of logs. 
\end{enumerate}

\subsection{Defining Sets}
\label{sec:defset}
To define the metric space, we first define our set $X$, by transforming the logs into a standard comparable representation (steps (1) and (2)). 
To do so, we adopt database terminology and methodology, transforming a non-relational structure of nested keys and values (e.g. Fig. \ref{fig: eventxml}) into a uniform relational structure (e.g. Fig. \ref{fig:cryptotempfilelogs}). 

The first step is to simply un-nest the levels of keys and values into a flattened list, appending a number to any keys that result in a non-unique key. 

Next, we chose a small subset of $N$ keys (or attributes) that we deem  useful, and the appropriate domain (value type) for the values of each key. 
Note that if a record does not have any of the $N$ keys in our subset, we discard it. 
We allow three possible  domains: numerical, string, or categorical, where a categorical attribute with $n$ possible categories is encoded as a length $n$ one-hot vector, a vector of all zeros except at the index of the category it represents, denoted $\mathbf{e_i}$. 
With this choice of attributes and domains we essentially define a relation schema for our log data. 
Formally, let $A_1, A_2, ...,A_N$ be the attribute names with the associated attribute domain sets $X_1, X_2, ...,X_N$. 
We define the relation schema $R$ for our data as $R(T, A_1:X_1, A_2:X_, ... A_N:X_N)$, where $T$ is time (specifically, UNIX time i.e. seconds since 01/01/1970).  
We define a \textit{log entry}, $x$, as an $n$-tuple in our relation, that is, $x = (t, x_1,..., x_N) \in  T \times X_1 \times ...\times X_N$. 

Additionally, since logs are comprised of a non-uniform set of attributes, we must have a way of dealing with ``missing'' data. 
If a log does not contain a particular attribute, we encode it with the the special null element for that attribute type (see Table \ref{table:metricdef}). 
Thus, we have now made our logs comparable by reducing each to be represented by a uniform set of attributes, as well as reducing the dimensionality and number of logs of our data. 
Furthermore, formulating our metric space using these general types allows us to apply the same procedure regardless of the subset of attributes that we choose, making our method both flexible and configurable. 

Shown in  Fig. \ref{fig:logentries} are two log entries in the following chosen schema: \textit{Entry(Time: numerical, BaseFileName: categorical, BaseFileExtn: categorical, CreatorProc: categorical, TokenElevation: numerical, CmdLine: string)}. These are consecutive log entries (30s apart) from the same IP during the beginning of a Cryptowall attack. Note that these are the transformed versions of the rightmost logs in Figures \ref{fig:cryptotempfilelogs} and \ref{fig:explorerlogs}.  
In the log on the right we added a null entry for the \texttt{CmdLine} attribute. Note that we derived categorical \texttt{BaseFileName} and \texttt{BaseFileExtn} attributes from the basename and extension of the \texttt{NewProc} path.  
This example highlights another way in which this method is highly configurable\textemdash by breaking attributes into different domains, emphasis can be put on different aspects of the data.  
\begin{figure}[ht]
\centering
\noindent\begin{minipage}{.24\textwidth}
\begin{lstlisting}[language=json]
{ Time (Unix): 1446532736
  BaseFileName: 4328,
  BaseFileExtn: TMP,
  CmdLine: "C:\Users\hostX\AppData\Local\Temp\Low\4328.tmp"
  CreatorProc: IEXPLORE,
  TokenElevation: 3 }
\end{lstlisting}
\end{minipage}
\begin{minipage}{.24\textwidth}
\begin{lstlisting}[language=json]
{ Time (Unix): 1446532766,
  BaseFileName: EXPLORER,
  BaseFileExtn: EXE,
  CmdLine: "",
  CreatorProc: 4328,
  TokenElevation: 3 }
\end{lstlisting}
\end{minipage}
\caption{Consecutive log entries (30s apart) from the same IP with five chosen attributes and time: categorical attributes are capitalized and string attributes are shown with quotes. Left: transformed version of Fig. \ref{fig:cryptotempfilelogs} (right). Right: transformed version of Fig. \ref{fig:explorerlogs} (right). }
\label{fig:logentries}
\end{figure}

Finally, we define a set for \textit{log streams}, time-ordered collections of log entries produced by a single IP. 
Formally, we define a \textit{log stream} $x(t)$ as an ordered set of $n$ log entries with order induced by time: $x: T \mapsto X_1 \times ... \times X_N $, denoted $x(t) = (x_1(t),...,x_n(t))$ with $x_j \in X_j$. We assume the host field is constant.

\subsection{Defining Metrics}
\label{sec:defmetric}
In this section, we tackle steps  (3) define a metric for each attribute type, (4) extend these to a metric on log entries, and (5) extend this to a metric on log streams.

\subsubsection*{Step (3) Attribute Metric Space}
\label{subsec:attributemetric}
We define the three general metric spaces for each of the attribute types in Table~\ref{table:metricdef}. We define metrics $d_i$ on each attribute type such that $d_i:X_i \times X_i \mapsto [0,1]$.
\begin{itemize}[wide=\parindent]
\item[\emph{Categorical:}]
We define the metric on categorical attributes as the normalized  $\ell^1$ distance.
\item[\emph{Numerical:}] We define the metric on numerical attributes as the normalized absolute difference between two reals.
\item [\emph{String:}] We define the metric on string attributes as a normalized Levenshtein distance. The Generalized Levenshtein Distance (GLD) is an edit distance which is defined as the minimum cost of transforming string $x_1$ into $x_2$ where each edit operations (insertion, deletion, replacement) is associated with a particular cost. In our implementation, we weight each of these edit operations equally with cost 1. We can therefore define the GLD between two strings as:
\begin{equation}
    \label{eq:1}
  d_{lev}(x_1, x_2) = \min\left\{{O(x_1, x_2)}\right\} 
\end{equation}
where $O_{x_1,x_2} = O_1O_2....O_l$ is a sequence of $l$ elementary edit operations used to transform $x_1$ into $x_2$. 
However, $d_{lev}(x_1, x_2) \in [0, \max\left \{ |x_1|,|x_2| \right \}]$ is not a metric in [0,1] and the obvious solution of dividing the result by the length of the longer string results in a violation of the triangle inequality \ref{defn:metricspace}. To solve this, we adopt a simple normalized GLD metric proposed by Yujian et. al.~\cite{yujian2007normalized}. 
See Table \ref{table:metricdef}. 
\end{itemize}

\begin{table}[t]
\begin{center}
\setlength\tabcolsep{0pt}
    \setlength{\extrarowheight}{7pt}
\begin{threeparttable}
\caption{Attribute Metric Space Definitions}
\label{table:metricdef}
\begin{tabular}{lccc}
    \toprule
    \textbf{} & \textbf{Set element} & \textbf{Metric} & \textbf{Null element} \\ 
    \midrule
      Categorical\tnote{1} \phantom {AA}  & 
      $\mathbf{e_i} \in \mathbb{R}^{N+1}$ & $\frac{\norm{\mathbf{e_i}-\mathbf{e_j}}_1}{2}$  & 
      $\mathbf{e_1} \in \mathbb{R}^{N+1}$\\
      Numerical   &
      $x_i \in \mathbb{R}$ &
      $\frac{|x_i-x_j|}{x_{\max}-x_{\min}+1}$ & 
       0 \\
      String\tnote{2}  &
      \phantom{AA} $x_i = \texttt{string}$  \phantom {AA} &
      $\frac{2  d_{lev}(x_i,x_j)} {|x_i|+|x_j|+d_{lev}(x_i,x_j)}$ &
      ``''\\
      \bottomrule
\end{tabular}
\begin{tablenotes}
\item[1] $N$ = number of categories \item[2] $d_{lev}$ defined Eq. \ref{eq:1}
\end{tablenotes}
\end{threeparttable}
\end{center}

\end{table}

\subsubsection*{Step (4) Log Metric Space}
    
We define the metric between two log entries, $d_{entry}: (\times_j X_j)^2 \mapsto [0,\sqrt{N}]$, as the $\ell^2$ difference of their attribute distances: 
\begin{equation}
 \label{eq:2}
    d_{entry}(x^{(1)}, x^{(2)}) := \sqrt{\sum_{i=1}^{N}d_i^2(x^{(1)}_i,x^{(2)}_i)}
    \end{equation}
It is a straightforward to show that $d_{entry}$ satisfies Def. \ref{defn:metricspace}.

\subsubsection*{Step (5) Stream Metric Space} 
We define the distance $d_{stream}: \{x(t): T \mapsto \times_j X_j\}^2 \mapsto [0,1]$ 
between two streams $f(t), g(t)$ as the $\ell^1$ distance 
$$d_{stream}(f(t), g(t) ): = \int_t d_{entry}(f(t),g(t))dm(t).$$ 
If we consider $f,g$ as a sequence of log entries (respecting only the chronological order of logs), then $f=(f_1,...,f_n)$, $g = (g_1,..,g_n)$, with $f_i$ and $g_i$ as the $i$\textsuperscript{th} log in their respective streams, and $m$ is defined as normalized counting measure. Hence,
\begin{align}
\label{eq:3}
d_{stream}(f,g)  :=& \int_{i=1}^{n}d_{entry}(f(i),g(i))dm(i) \nonumber\\
            =& \frac{1}{n}\sum_{i=1}^{n}d_{entry}(f(i),g(i))\ .
\end{align}

On the other hand, if we wish to respect the specific times of the entries in the log streams $f, g$, we a priori have logs with different timestamps,  $f = (f(t_0),...,f(t_n))$, $g = (g(s_0),...g(s_n))$.
A benefit of this metric space embedding is that we can interpolate using convex combinations. Specifically, if $t_0 < s < t_1$, then $s = \lambda t_0 + (1-\lambda) t_1$ for some $0 < \lambda < 1 $, and we define  $f(s) := \lambda f(t_1) + (1 - \lambda )f(t_2)$. We can then define the distance over $f, g$ w.r.t time as:
\begin{equation*}
d_{stream}(f,g) :=
\frac{1}{t_n-t_0}\sum d_{entry}(f(u_i),g(u_i))
\end{equation*}
for a fixed interpolation of $\{u_1, ..., u_k\}$ of the interval $[t_0, t_n]$. 

Note that for a discontinuous stream, $f$, $d_{stream}(f(t), f(t+\epsilon))$  can be arbitrarily large. 
In practice one must take caution to align streams in the time domain to use this distance. 
See experiment in Sec. \ref{sec:malwaredection} where cross correlation is applied. 

Altogether this may seem at first glance overly formal, it has the advantage that the metric space calculations at all three levels can be coded and applied to any set of logs for a schema so long as the values are categorical, numerical, or string types.




\subsection{Requirements Revisited} 
In Sec. \ref{sec:requirements} we enumerated requirements for a desired metric on logs. Here, we evaluate our metric in terms of those requirements, and calculate the pairwise distances between our initial example logs provided in the requirements section. 

We note that by definition, our metric space of logs, and therefore streams, satisfies requirement (1), that the metric uses attribute-wise distance measures designed to preserve semantic similarity. 
As an example, using the attributes defined in the schema above, the distances between the two logs shown in Fig \ref{fig:cryptotempfilelogs} is about 1.011. On the other hand, the distance between the leftmost logs in Fig. \ref{fig:cryptotempfilelogs} and Fig. \ref{fig:explorerlogs}, which are much more dissimilar, is about 1.866 (note that the maximum possible distance for entries with five attributes is $\sqrt{5} \approx 2.236$).

Requirement (2), that we can compute averages / cluster centers is more nuanced. 
Since numerical and categorical attributes are embedded into a normed vector space, averages / centroids are straightforward\textemdash sums and scalar multiplication are well-defined and by definition these operations are respected by the norm. 
In short, if your schema (chosen set of keys from step (2)) does not include strings Requirement (2) is satisfied. 
For strings, there is no straightforward embedding to a vector space that preserves similarity, so defining an actual centroid is not straightforward. 
On the other hand, defining the distance to a theoretical centroid gives a workaround.  Specifically, given a multiset of strings, $S = \{s_1, ..., s_k\}$, we define the distance from a string $s$ to the ``centroid'' of $S$ as the average of $ d_{str}(s,s_i)$. If we take the centroid of all of the new windows explorer process logs seen in our dataset, and compute the distance from this centroid for each of the logs in Fig. \ref{fig:explorerlogs}, we get the distances (from left to right): 0.5, .625, .875. These can be seen as a measure of how unusual each log is, so using this metric, the malicious log is the most unusual.  

Requirement (3) is that the metric takes into account the sequence of logs, which is true by definition for the metric on log streams. 
For an example, see Sec. \ref{sec:malwaredection}, where the sequence of logs for an observed attack is used as a ``soft signature'' to identify later polymorphic occurrences.

%% file: 40-experiments.tex
\section{Experiments \& Results}
\label{sec:experiments} 
In this section we apply this approach to two datasets in two experiments that exemplify the utility of defining a metric space for Windows system logs. We emphasize that following experiments are only possible due to the generalized embedding we rigorously defined in \ref{sec:metricspace}.
Because of the quantity of data used, 
these experiments should be treated as example use cases rather than  well-tested experiments.  

In all of our experiments we use the following schema (sometimes dropping \textit{CmdLine}): 
\emph{Entry(Time, BaseFileName: categorical,BaseFileExtn: categorical, CreatorProc: categorical, TokenElevation: numerical, CmdLine: string)}. 
These particular attributes are generally what characterize a New Process Log, specifically, a log of type \texttt{Security} with the EventID \texttt{4688}. 
So, in forming our set of log entries as described in Sec. \ref{sec:defset}, and including only logs that contain at least one of these attributes, we are essentially subsetting to include only New Process logs, which makes up approximately 84\% of our data. 

\subsection{Pre-encryption Conviction of Polymorphic Ransomware}
\label{sec:malwaredection}
The goal of this experiment is to show how the metric space permits organic creation of ``soft signatures'' to detect ransomware from the usual log collection data. 
By ``soft signature'' we mean a behavioral signature that provides a continuous distance-driven score that in theory will detect polymorphic samples. 

Ransomware generally executes a sequence of events before  encrypting the users' data, e.g. dropping files, unpacking itself, deleting shadow copies of files in memory, beaconing to a command and control server, exchanging encryption keys \cite{chen2017automated}; hence, the goal for defense against ransomware is pre-encryption conviction (aside from whitelists, firewalls, and other standard security postures). 
We consider the following scenario\textemdash when a first computer at an organization is infected and encrypted by a novel ransomware, can the SOC create a polymorphicly-robust, pre-encryption detector simply from the host logs collected (that is, without malware samples to analyze)? 

\subsubsection*{Data} 

For the experiment, we use actual logs containing Cryptowall 3.0, a polymorphic ransomware. We take a sequence of logs during the attack and compare it to sequences of logs from equal length segments of test data using the log stream distance. This dataset consists of seven streams from six different IPs collected in the same week. 
Four of these are `attack streams' from different IPs that contain a Cryptowall 3.0 attack. In each of the instances, logs for leading up to execution of Cryptowall appear, and the ambient logs are mixed with the ransomware's actions. In all cases the trojan is eventually caught and quarantined by antivirus software before encryption. The other three are ambient streams, one of which is from the same IP (host 4) as an attack stream, but from the previous day.

\subsubsection*{Experimental Design}

We consider one attack stream as the initial attack, and use only this to build a soft signature. 
First we extract the {\it observed attack sequence}, denoted $f_{sig}$, the sequence of logs from the initial attack vector log to right before the malware contacts the home server to request a key for encryption. 
(This requires manual analysis of the log stream to truncate it.) 
In each of the four instances of attacks in our data the attack sequences, $f^{(1)}_{sig}, ..., f^{(4)}_{sig}$ range from six to seven log entries and the time intervals range from 36 to 44 seconds. 
It is worth noting that these signatures do not contain any logs created by a malware or virus protection software, and they occur before the encryption has begun.
Additionally, the attack sequences, $\{f^{(i)}_{sig}\}$ differ not only due to the polymorphic nature of the attack, but also due to the presence of ambient logs unrelated to the malware. 

Our proposed technique is to build an online detector, comparing the attack sequence from the first infection, $f_{sig}$, to the recent log stream from a host, and if the distance between the two streams are sufficiently close, alert. 
In practice (we envision that) upon receipt of a new log, $l_k$, the distance  from $f_{sig}$ to the previous $n_{sig}:= |f_{sig}|$ logs is computed, $d_{stream}(f_{sig}, [l_{k-n}, ..., l_k])$ and compared to the previously observed distances to determine if it is anomalously low---indicating the most recent sequence of logs for that host is similar to first attack sequence observed. 

\begin{figure}[t]
    \centering
    \includegraphics[width = .45\textwidth]{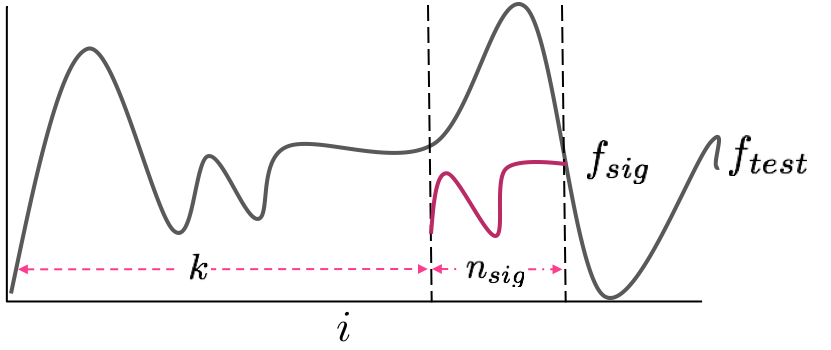}
    \caption{Representation of cross correlation between a test stream $f_{test}$, and an observed sttack sequence $f_{sig}$.}
    \label{fig:crosscorr}
\end{figure}

To test this hypothesis experimentally with our data,  we use cross-correlation to quantify the degree of similarity between a fixed attack sequence $f_{sig}$, and each of the log sequences in our data set. 

While cross correlation usually refers to the displaced dot product between two signals, here we use it to refer to the displaced stream distance between two signals.  
Specifically, with fixed $f_{sig}$ as above, 
and a test sequence $f_{test}$, we compute the distance $d_k$ between each $n_{sig}$-length sub-sequence of $T$ (Fig. \ref{fig:crosscorr}):
\begin{align*}
d_k :   =& Corr \{f_{sig},f_{test}\}[k]\\
        =& \sum_{i}^{n_{sig}} d_{entry}(f_{sig}[i],f_{test}[i+k]). 
\end{align*}
over each $k= 0,..., n_{test}-n_{sig}$ displacement in the test stream.
Hence, we obtain a sequence of distances from $f_{test}$, $d_1, ..., d_{l}$ where $l = n_{test}-n_{sig}$. 

In order to build an online detector, 
we give an extreme value score to each distance observation, $d_k$, defined  as  the probability of seeing a smaller distance. 
To compute this probability, we use a p-value of a $t$-distribution, fit to the observations we have seen up to that point, i.e., $d_1, \dots, d_{k}$. 
We choose the $t$-distribution due to the small sample size.

By computing the extreme value scores in this way, we are able to simulate how this measure could be used as a real time / online  detector. 
We compute the test statistic for $d_k$ where the parameters are estimated from the previously seen distances,
\begin{equation*}
t^*_k = \frac{d_k - \overline{D_k}}{s_k}
\end{equation*}
where $D_k = \{d_1,...,d_k\}$, $\overline{D_k}$ is the sample mean of $D_k$, and $s_k$ is the sample standard deviation of $D_k$. We then compute the one sided p-value by calculating the tail probability of each data point from the CDF of the t-distribution with $k-1$ degrees of freedom, i.e.,  p-value($d_k)= \mathbb{P}(T <t^*_k)$. 

\subsubsection*{Results}
The minimum p-values for each attack signature against every full stream of logs is shown in Table \ref{table: cryptowallresults}. 
Our results indicate that the attack intervals are at least 11 orders of magnitude closer to the generated signatures than ambient intervals, where this minimum p-value for the attack streams does in fact correspond to the attack interval.
While this extremely small distance is unremarkable when comparing the attack signature to the stream from which was was extracted, it is clear that a signature generated from a different attack also produces an extremely low p-value. 
We recall that the four occurrences of Cryptowall in our data experienced polymorphism, specifically, using different file names across samples. 

Given the insufficient amount of attack data, we cannot say with certainty whether this approach would lead to an accurate detector, but these results indicate the merits of using this metric for detecting malware events.
If for example, we were to create detector by setting a p-value threshold of 1E-15, and test each signature against all the data streams (not including the one it was drawn from). We get no false positives, and detect the attack in an average of 40.24 seconds, directly before the sample contacts the home server to request a key for encryption. 

This approach promises multiple improvements over state-of-the art commercial AV software that was deployed on the hosts in the experiment. 
First, it promises to identify the ransomware long before it was actually detected by standard AV software running on these hosts. As a comparison, the software running on the hosts in our examples took between 1-4.5 minutes to quarantine the ransomware---in two out of four of the cases, our method detects more than four times faster.
Secondly, it promises detection that is robust to polymorphism because it uses similarities across all chosen keys and across time-varying sequences of logs (corresponding to malware behaviors). 
Third, it does not require  a sample of malware to craft signatures; rather once an attack is identified, the attack sequence from ambient logs theoretically be shared and used for the detectors ``soft signature''. 
Finally, manual analysis of malware is not needed, although truncating the attack sequence of the logs is needed. 

\begin{table}[t]
\begin{center}
\begin{threeparttable}
\caption{Log stream for IPs 1-4 included a Cryptowall attack; IP 4 included a day without an attack; IPs 5,6 had no attack. 
For each attack, the attack sequence of logs, $f^{(i)}_{sig}$ is slid across each full sequence of data to obtain a sequence of stream distances. 
P-values from $t$-distribution used for determining if each subsequence is extremely close to attack sequence. 
Minimum p-values reported.  
Notice that sequences containing an attack have at least one subsequence that is at least 11 orders of magnitude smaller than for non-attack sequences. 
This shows that observing Cryptowall in a single log can create an accurate detector for all other logs including polymorphic, identical, or no Cryptowall attacks. }
\label{table: cryptowallresults}
\begin{tabular*}{0.5\textwidth}{cccccc}
\toprule
& IP & $f^{(1)}_{sig}$ & $f^{(2)}_{sig}$    & $f^{(3)}_{sig}$    & $f^{(4)}_{sig}$  \\\midrule
\multirow{4}{*}{Attack}
& 1  & 1.45E-25 & 4.23E-21 & 4.34E-21 & 9.70E-17 \\ 
                            & 2  & 1.79E-21 & 5.19E-44 & 1.12E-34 & 3.44E-22 \\ 
                            & 3  & 7.78E-24 & 6.49E-58 & 1.08E-82 & 1.62E-30 \\ 
                            & 4  & 6.39E-21 & 1.83E-31 & 1.92E-31 & 1.94E-67 \\
    \midrule
\multirow{3}{*}{Non-Attack} & 4  & 4.87E-09 & 3.39E-09 & 3.51E-09 & 1.29E-05 \\ 
                            & 5  & 1.79E-04 & 1.29E-03 & 1.30E-03 & 5.81E-04 \\ 
                            & 6  & 1.95E-05 & 2.60E-05 & 2.81E-05 & 9.52E-05 \\ \hline
\end{tabular*}
\end{threeparttable}
\end{center}
\end{table}

\subsection{IP Classification}
\label{sec:ipclassification}

The goal of this experiment was to validate the theory that similar user activity clusters in the metric space we defined. 
Following time-series analysis techniques in \cite{bridges2018towards}, we reduce the dimensionality of the dataset using the log entry metric, and then attempt to classify data streams by IP and user role, as well as identify outlying streams that indicate a deviation from a users' normal behavior. 


\subsubsection*{Data} We use log data from seven different IPs collected over a period of five to ten days. These seven users fall into three general roles: 3 administrators, 1 IT, and  3 researchers. For the purposes of these experiments, we split the data from each IP by day, and represent each `daily stream' $f$ as as a set (i.e. $f=\{f_1,...,f_n\}$ is an unordered set of logs collected from a single IP over the course of one day). In this case, we opt for this simple  structure that does not respect the chronological nature of the data because we are interested in comparing overall behavior, not in examining events. 
Additionally, in this experiment we chose to not use the \textit{CmdLine} attribute.
    
\subsubsection*{Experimental Design} We first take the centroid of each of the 61 streams. Each centroid represents the average activity for an IP over one day. We then compute the pairwise distance between each stream using the log entry metric (Eq. \ref{eq:2}), creating a $61 \times 61$ distance matrix. Using this distance matrix, we: (1) use a dimensionalty reduction technique to visualize the distance between streams, and (2) use a $K$-Nearest-Neighbors (KNN) classifier to classify the streams in terms of user IP and role. 

In order to visualize the similarities of these streams, we use metric Multi-Dimensional Scaling (mMDS) \cite{borg2005modern} to find a two dimensional representation of the data. mMDS is a method of finding a lower dimensional representation that preserves the notion of similarity provided by a metric in a higher dimensional space.

\subsubsection*{Results}

%

\begin{figure}[t]
\centering
\includegraphics[width=0.48\textwidth]{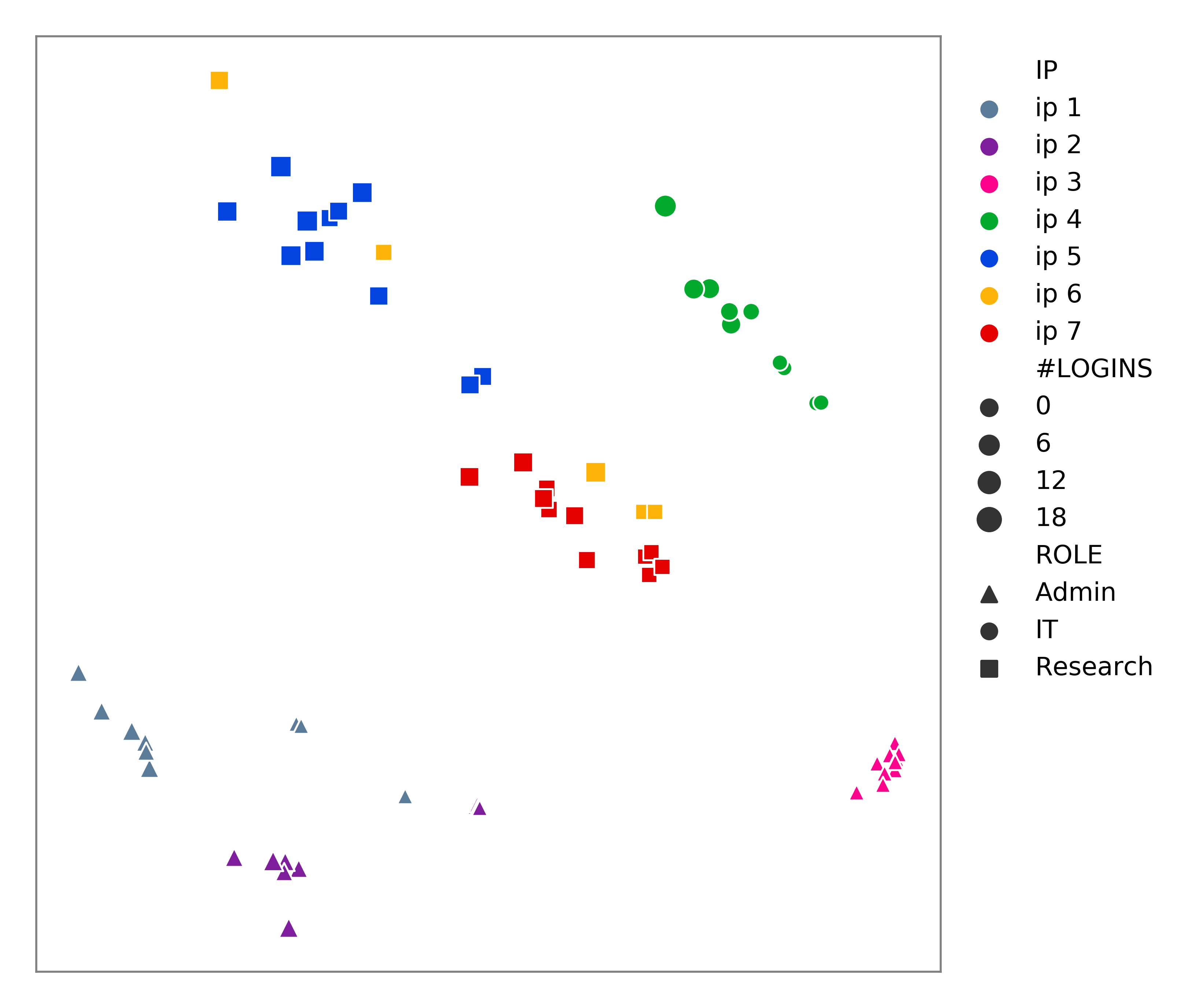}
\caption{Daily log streams by user IP and role, visualized using mMDS dimensionality reduction technique. Streams from the same IP seem to form dense clusters (with the exception of IP 6), and clusters from IPs with the same user role seem to be closer together. Days in which a user did not log in (illustrated with the smallest points) form particularly dense clusters. Using $K$-Nearest-Neighbors, classifying by role yields perfect results, and classifying by IP, we get very high scores with the exception of IP 6 (see Table \ref{table:ipclassification}).}
\label{fig: ip_role_viz}
\end{figure}

We can see the results of the visualization using mMDS in Fig. \ref{fig: ip_role_viz}. It is clear that streams from the same IP for the most part cluster quite densely, and that individual administrators in particular tend to have similar activity across different days. One exception to this appears to be IP 6, which has two anomalous streams in the second quadrant, in which they seemed to deviate from their normal daily activity.
We can also see that streams from IPs with the same user role tend to cluster closely together. 
Even in the case of the outlying streams for IP 6, these still cluster quite closely to the streams of another researcher (IP 5), perhaps indicating that these days were not out of character for a researcher in general. 
Finally, we see that days in which there were few or no log ins (i.e. stream is entirely background processes), tend to cluster more densely slightly apart from other other higher activity streams from the same IP, and are generally closer to the fourth quadrant. 
This seems to indicate that this method is at least partially picking up on desktop configurations rather than user behavior. 
It is worth noting that although this is visualized in a two dimensional space, these observations are still relevant in our higher dimensional representations. 
For example, if we take the variance over each set of IP streams using our log entry metric, we find that IP 3 has the lowest variance (.0014), IP 6 has the highest variance (.071) and that the rest range between .018 and .037, which tallies with our visualization.

We now perform KNN classification using Leave-One-Out Cross Validation (in which all streams for a particular IP are held out when classifying user roles). 
Choosing $K=3$ yielded the highest accuracy. 
Classifying by user role, we obtain a perfect classification, although we are unable to get results for the IT role due to the fact that we only have an example from one IP.   
When classifying by IP we obtain slightly worse results, particularly for IP 6, and the resulting average F1 scores are shown in Table \ref{table:ipclassification}. 
These very high scores illustrate that this method is effective in classifying users by role, and to a lesser extent, for classifying users by IP.

\begin{table}[t]
    \centering
    \caption{Leave-One-Out Cross Validation F1 Scores for User IP Classification (KNN $K=3$): KNN Classification yields high scores for all IPs except for IP 6, which agrees with visual inspection of Fig. \ref{fig: ip_role_viz}. This shows that an IPs behavior across multiple days tends to be similar in our defined metric space. Note that incorrect labels were misclassified as IPs from the same user role, and classification by role yielded perfect results.}    \label{table:ipclassification}
    \begin{tabular}{ccccccccc}
    \toprule
           IP & 1 & 2 & 3 & 4 & 5 & 6 & 7 \\
            F1-Score & 0.952           & 1.0          &  .941         & 1.0         & 0.917     & 0.75 & 1.0      \\\bottomrule
    \end{tabular}
\end{table}
        

However, when we repeated the same procedure with streams that contained no logins, and are therefore unrelated to user behavior, we obtained only slightly worse results, leading us to believe that much of what this method was picking up on desktop configurations rather than user behavior. 

This highlights one particularly problematic aspect of working with Event Log data, namely, logs that are produced by background processes, and are not of interest in examining user behavior, can vary significantly across IPs.
If there are disproportionately high number of background logs, this noise tends to overshadow any distinction one could hope to see in those produced by user-initiated activity. 

Despite this finding, this method is still promising for comparing streams produced by the same IP, and could be used as a tool for pinpointing  anomalous streams which might  warrant further investigation by an operations department, e.g.,  hunting insider threats. 
For example, a simple anomaly detector could be implemented by establishing a baseline variance for each IP, and flagging any stream that deviates more than a given number of standard deviations as suspicious.

\subsection{Visualizing/Characterizing Daily Activity}
\label{sec:vizactivity}

This final experiment is focused on unsupervised exploratory visualizations that could potentially be used by a SOC operator to get an overview of the daily activities of the hosts on their network. 
As discussed above, a log stream is generally dominated by background processes, and these tend to wash out signals produced by user-initiated activity. 
However, by applying clustering techniques, we can in fact use the overwhelming volumes, consistency, and repetitive nature of these background logs to our advantage. 
Specifically, we perform two different methods of clustering to provide meaningful dimension reduction to inform novel visualizations. 
In the first, to understand hourly trends per IP, we use bagging features of logs for each hour, an intermediate representation between respecting the sequential relationship of logs (as in Sec. \ref{sec:malwaredection}) and bagging features (neglecting chronology as in Sec. \ref{sec:ipclassification}).  
In the second to understand sequences of events for an IP, we respect the chronological occurrence of logs for short sequences. 
These visualizations demonstrate how this method could be used to decompose log streams from co-occurring processes, identify recurring background processes, and establish a baseline for a users daily activity.

 
\subsubsection*{Data} We use the same data collected from seven IPs from three different roles as in Sec. \ref{sec:ipclassification}, and again omit the \textit{CmdLine} attribute.
    
\subsubsection*{Experimental Design (A)} First, we demonstrate how our metric facilitates the application of the $k$-means clusering algorithm, and how this can quickly and easily be used to to get an overview of the behavior of various IPs on a network. 
We initially split each stream into hour-long segments, and take the centroid of each substream, thus representing each hour for each IP with a single aggregate log.
We then bin each IP-hour into $k$ bins using $k$-means clustering with the log entry metric (Eq. \ref{eq:2}). We randomly choose initial ``seed'' centroids by using a method similar to $k$-means++, in which the probability of choosing each successive centroid is inversely proportional to its distance to the closest existing seed. 
We select the parameter $k$ through the elbow method, plotting the within-cluster sum of squares as a function of number of clusters over $k \in \{2, ..., 10\}$. 
We find that the marginal decrease in unexplained variance drops after $k=3$ and again at $k=5$, and we choose $k$ to be 5. 
The clusters represent the most dominant processes for this group of IPs, and we can visualize similar reoccurring  processes across IPs.

\subsubsection*{Results (A)}

\begin{figure}[t]
\centering
\noindent\begin{minipage}{.23\textwidth}
\includegraphics[width = \textwidth]{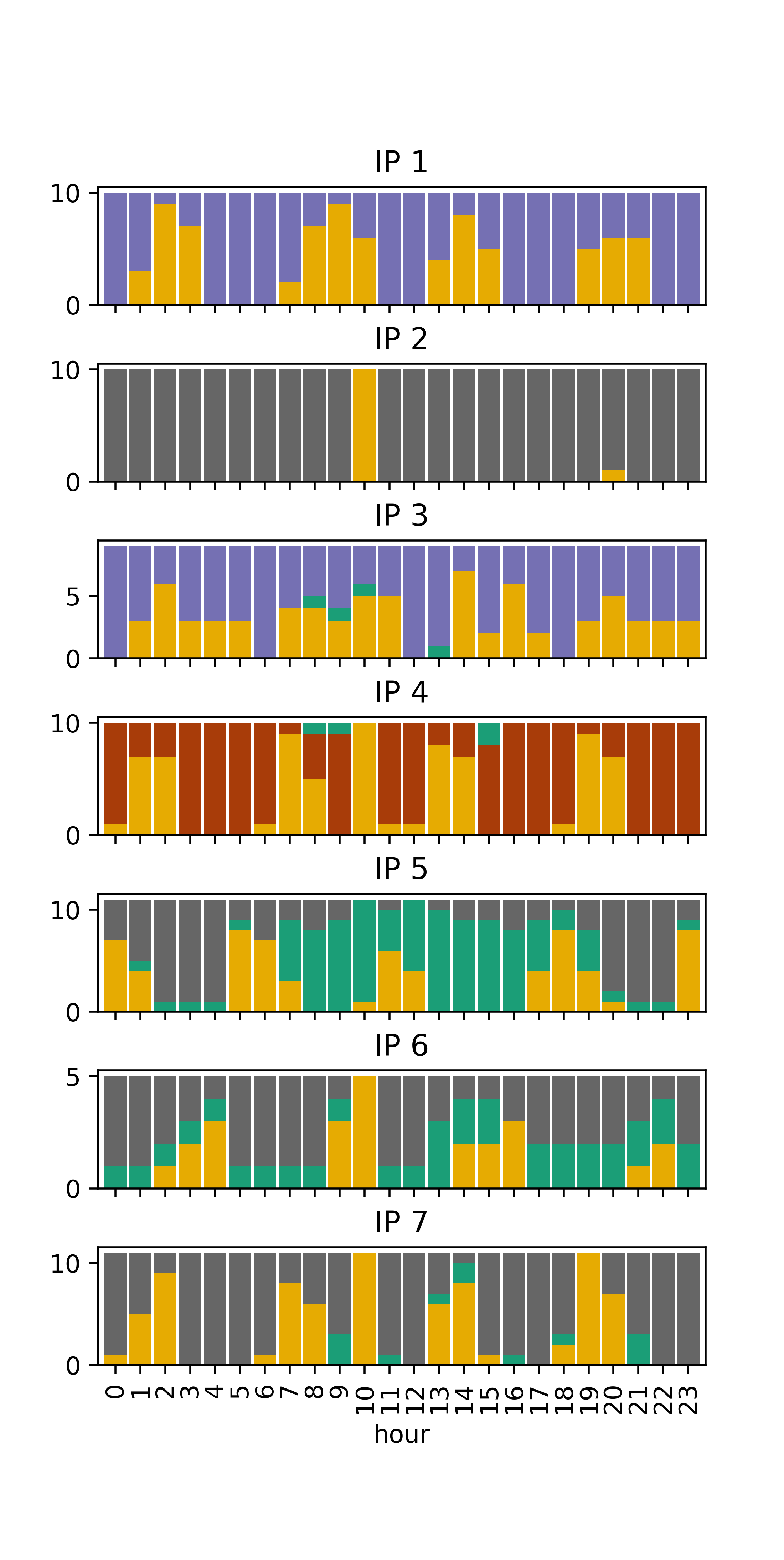}
\end{minipage}
\begin{minipage}{.25\textwidth}
\caption{Histograms depicting the daily activity of each IP over the course of five to ten days. Colors indicate cluster label (i.e. of the ten days of data for IP1, nine of 2AM-3AM windows fell into the yellow cluster). Centroids show that the \textbf{{\color{Dandelion} yellow}} cluster is comprised of a Microsoft background controller service that is set to run several times a day, and the (\textbf{{\color{PineGreen} green}}) cluster is made up mostly of Google Chrome background processes. These results support the theory that much of the distinction seen in Fig. \ref{fig: ip_role_viz} are due to background processes: main background processes are different for admins (IP1, IP3: \textbf{{\color{Periwinkle} purple}}), researchers (IP5-7: \textbf{{\color[rgb]{.5,.5,.5} gray}}), and IT (IP4: \textbf{{\color{Maroon} Brown}}).
Additionally, it is clearer why IP2 differs so significantly from the other admins\textemdash the Microsoft process occurs only once a day at 10AM.}
\label{fig: ip_day_hist}
\end{minipage}
\end{figure}

We show the results from the $k$-means clustering of IP-hours in a histogram in Fig. \ref{fig: ip_day_hist}. 
This shows the frequency of each label seen for each IP over the five- to ten-day collection period. Visualizing the data in this way allows us to easily identify reoccurring processes across days and IPs. For example, the process illustrated in {\color{Dandelion} yellow} appears to occur roughly four times a day for all of the IPs with the exception of IP 2. By examining the centroid of this cluster, we find that this is an Microsoft background controller service that runs periodically and begins with a new `policyHost.exe' process. Comparing this to Fig. \ref{fig: ip_role_viz}, it becomes apparent that much of the distinction found between IPs and roles that was likely due to background processes.

\subsubsection*{Experimental Design  (B)} Second, we show an alternative method of clustering that leverages the sequential aspect of the data.  
We split each stream into short, $m$-length sub-sequences, compute the pairwise stream distance between each using the sequence stream metric (Eq.\ref{eq:3}). 
We then perform DBSCAN \cite{ester96adensity}, a density-based clustering algorithm, to group similar sub sequences into clusters and identify outlying sub-sequences. 
Due to the high volume of background process logs, we hope to label similar background events (clusters), and identify user initiated processed (outliers). 
We used $m = 7.$

Note that in both of these experiments, since we have designed our metric to preserve semantics, we can easily understand the meaning of these clusters by computing and examining cluster centroids. For the categorical attributes, the vector can be interpreted as the percentage of each category in the cluster (see bottom Fig. \ref{fig: clustered_day_stream} showing examples of the centroids of two Microsoft background process sequences).


\subsection*{Results (B)}

\begin{figure}[t]
\centering
\fbox{\includegraphics[width=.47\textwidth]{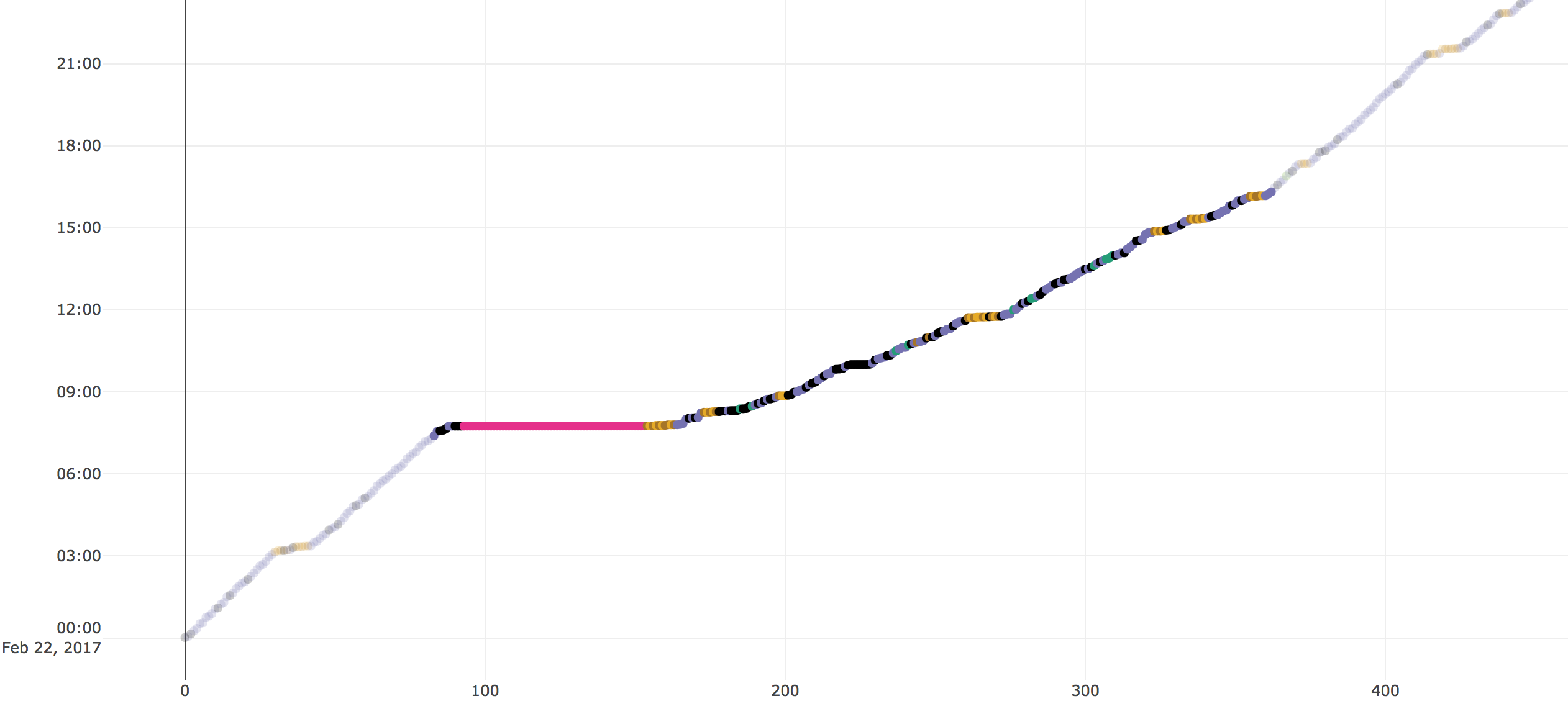}}
 \hspace{1.0cm}
\fbox{\includegraphics[width=.47\textwidth]{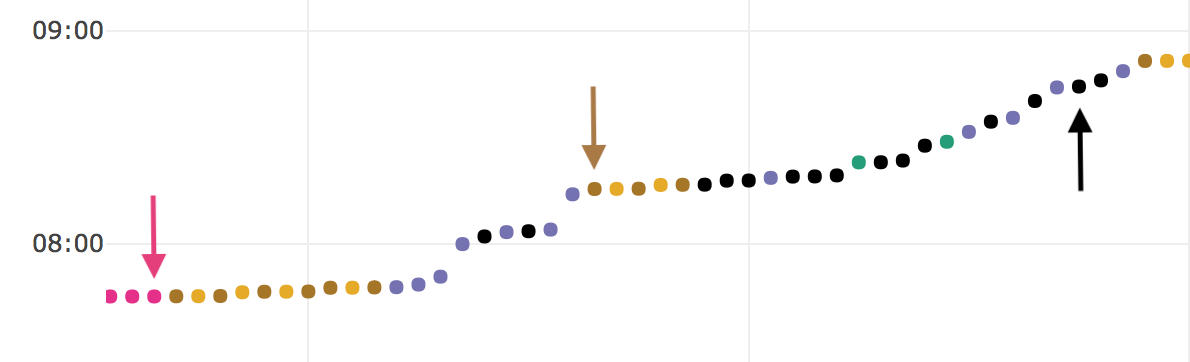}}
\noindent\begin{minipage}{.47\textwidth}
\noindent\begin{minipage}{.24\textwidth}

\begin{lstlisting}[language=json, frame=lines, rulecolor=\color{Magenta}, showspaces=false]
CreatorProc:
 [NGENTASK: 0.86
  NGEN: 0.14]
TokenElevation: 1.0
BaseFileName:
 [NGEN: 0.86
  MSCRSVW: 0.14]
BaseFileExtn:
 [EXE : 1.0]

\end{lstlisting}
\end{minipage}\hfill
\begin{minipage}{.32\textwidth}
\begin{lstlisting}[language=json, frame=lines, rulecolor=\color{Dandelion}, showspaces=false]
CreatorProc:
 [POWERSHELL: 0.43
  CSC: 0.29
  WMIPRVSE: 0.29,
  SERVICES: 0.14]
TokenElevation: 1.14
BaseFileName:
 [CSC: 0.29
  CVTRES: 0.29
  CONHOST: 0.14
  POLICYHOST: 0.14
  POWERSHELL: 0.14]
BaseFileExtn:
 [EXE : 1.0]
\end{lstlisting}
\end{minipage}\hfill
\begin{minipage}{.32\textwidth}
\begin{lstlisting}[language=json, frame=lines, rulecolor=\color{black}, showspaces=false]
CreatorProc:
 [ACROCEF: 0.29
  ACROBAT: 0.29
  ADOBE_LICUTIL: 0.14
  EXPLORER: 0.14
  SVCHOST: 0.14]
TokenElevation: 3.0
BaseFileName:
 [ACROCEF: 0.43
  ACROBAT: 0.14
  ADOBE_LICUTIL: 0.14
  CONHOST: 0.14
  DLLHOST: 0.14]
BaseFileExtn:
 [EXE : 1.0]
\end{lstlisting}
\end{minipage}\hfill
\end{minipage}
\caption{Representation of a daily log stream for IP 3 with each seven-log subsequence represented by a colored dot. Dots are plotted subsequence number vs. time, where the color indicates the cluster classified by DBSCAN (outliers illustrated in black). Top plot shows entire day with logged-in period in bold colors (which is also notable by the decreased slope due to a higher log density). Note the long login sequence shown in \textbf{{\color{Magenta} pink}} which is comprised of Microsoft .NET Framework processes precompiling assemblies over about two minutes. Middle plot shows a closeup of the 8AM-9AM window containing the end of the login sequence, followed by a Microsoft Policy Host sequence (\textbf{{\color{Dandelion} yellow}}). This sequence is interwoven with other general (\textbf{{\color{Periwinkle} purple}}) and Chrome-related (\textbf{{\color{PineGreen} green}}) background processes, as well as outlying user-initiated processes (\textbf{black}). Examples of background processes, as well as a user initiated Adobe related process sequence, are highlighted, and their centroids are shown in the bottom figure.
Note that these colors have been chosen to show the correspondence between the clusters found using this method and that in Fig \ref{fig: ip_day_hist}. }
\label{fig: clustered_day_stream}
\end{figure}

The results for clustering $m=7$ length  sequences using DBSCAN is shown for a particular daily stream for IP 3 in Fig. \ref{fig: clustered_day_stream}. 
Each substream is depicted with a dot colored according to cluster. 
Additionally, we compute the centroids of the set of $m=7$ logs that make up each sub sequence, and display two examples of clustered background processes: Microsoft .NET precompiling process initiated at login (left), and the Microsoft `Policy Host' sequence also seen in Fig. \ref{fig: ip_day_hist} (middle).
Unlike the first method, this can be used to extract user initiated processed from a stream saturated with a high volume of background process logs. 
The closeup in Fig. \ref{fig: clustered_day_stream} shows user-initiated sequences (labeled as outliers) interleaved with background processes. 
We highlight and show the centroid for a particular user-initiated sequence in which a PDF was opened from Windows Explorer and viewed in Adobe Acrobat.

While these examples and visualizations are very basic, they demonstrate the ways in which clustering can be easily applied once we have a metric and how this could be implemented as a tool for SOC operators to quickly sort and visualize large volumes of data. 
Additionally, this latter clustering method could potentially be used to extract an active user signal from background process noise, and this could be used as a preprocessing step for Sec. \ref{sec:ipclassification}.

%% file: 50-conclusion.tex
\section{Conclusions} 
\label{sec:Conclusion}

In this paper, we demonstrate the need for developing a semantically meaningful similarity measure on Windows Event Logs and illustrate the requirements and benefits of such a measure using examples from real logs. 
We rigorously define a metric space on log attributes, extend this to a metric on log entries, and then to sequences of logs, and explain how this metric space fulfills the requirements of our similarity measure. 
We then demonstrate the utility and flexibility of this embedding with three use cases: automated behavior-based signature generation and detection for cyber attacks, grouping hosts by software profiles, and exploring background events. 
While our use cases are initial results, the results are compelling and our goal is to show the efficacy of our metric space for a variety of host-log analytic applications. 


%% file: 90-acks.tex
\section*{Acknowledgements}
\label{sec:acks} 
Special thanks to Jason Laska and anonymous reviewers. This material is based on research sponsored by the U.S. Department of Homeland Security (DHS) under Grant Award Number 2009-ST-061-CI0001, DHS VACCINE Center under Award 2009-ST-061-CI0003, DHS homeland security related science, technology, engineering and mathematics (HS-STEM) internships, and Laboratory Directed Research and Development Program of Oak Ridge National Laboratory, managed by UT-Battelle, LLC, for the U. S. Department of Energy, contract DE-AC05-00OR22725.  
The views and conclusions contained herein are those of the authors and should not be interpreted as necessarily representing the official policies or endorsements, either expressed or implied, of the DHS.